\documentclass[onecollarge,natbib]{svjour2}
\bibpunct{[}{]}{;}{n}{}{,} % to get "[numbered]" references from natbib
\smartqed  % flush right qed marks, e.g. at end of proof
\usepackage{graphicx}
\usepackage{amsmath}
\usepackage{amsfonts}
\newcommand{\be}{\begin{equation}}
\newcommand{\ee}{\end{equation}}

\newcommand{\bea}{\begin{eqnarray}}
\newcommand{\eea}{\end{eqnarray}}

\newcommand{\bi}{\begin{itemize}}
\newcommand{\ei}{\end{itemize}}

\journalname{Few-Body systems}
\begin{document}

\title{Baryon-to-meson transition distribution amplitudes: formalism and models%\thanks{Grants or other notes
}


\author{B. Pire         \and
        K.~Semenov-Tian-Shansky  \and
         L.~Szymanowski%etc.
}


\institute{B. Pire \at
              Centre de physique th\'eorique, {\'E}cole Polytechnique, CNRS, Universit\'e Paris-Saclay, F-91128 Palaiseau, France
              \email{bernard.pire@polytechnique.edu}           %  \\
           \and
           K.~Semenov-Tian-Shansky \at
              Petersburg Nuclear Physics Institute, RU-188300, Gatchina, Russia
               \email{kirill.semenov@polytechnique.edu}
                 \and
          L.~Szymanowski \at
              National Centre for Nuclear Research (NCBJ), PL-00-681 Warsaw, Poland
  \email{Lech.Szymanowski@ncbj.gov.pl}}

\date{Received: date / Accepted: date}

\maketitle

\begin{abstract}
In specific kinematics, hard exclusive amplitudes may be  factorized  into a short distance dominated part computable in a
perturbative way on the one hand, and universal, confinement related hadronic matrix elements on the other hand.
The extension of this description to processes such
as backward  meson electroproduction and forward meson production in antiproton-nucleon scattering leads to define new hadronic matrix elements of three quark operators on the light cone,
the nucleon-to-meson transition distribution amplitudes,   which shed a new light on the nucleon structure.
\end{abstract}

\section{Introduction}
\label{intro}
Accessing  transition distribution amplitudes (TDAs)
in specific exclusive reactions  is an important goal to
progress in our understanding of quark and gluon confinement.
The baryon-to-meson  TDAs are defined through baryon-meson matrix
elements of the nonlocal three quark (antiquark) operators on the light cone 
($n^2=0$):
\begin{equation}
\hat{O}^{\alpha \beta \gamma}_{\rho \tau \chi}( \lambda_1 n,\, \lambda_2 n, \, \lambda_3 n)
 =
\varepsilon_{c_{1} c_{2} c_{3}}
\Psi^{c_1 \alpha}_\rho(\lambda_1 n)
%[\lambda_1 n;\lambda_0 n ]
\Psi^{c_2 \beta}_\tau(\lambda_2 n)
%[\lambda_2 n;\lambda_0 n ]
\Psi^{c_3 \gamma}_\chi (\lambda_3 n),
%[\lambda_3 n;\lambda_0 n ],
\label{operators}
\end{equation}
where
$\alpha$, $\beta$, $\gamma$
stand for the quark (antiquark) flavor indices and
$\rho$, $\tau$, $\chi$
denote the Dirac spinor indices. Antisymmetrization is performed over the
color group indices
$c_{1,2,3}$.
Gauge links in
(\ref{operators}) are omitted by adopting the light-like gauge
$A \cdot n=0$.
These non-perturbative objects, first studied in
\cite{Frankfurt:1999fp, Pire:2004ie,Pire:2005ax},
share common features both with baryon distribution amplitudes (DAs) introduced in \cite{Lepage:1980,Chernyak:1983ej} as baryon-to-vacuum matrix elements of the same operators
(\ref{operators})
and with generalized parton distributions (GPDs), since the matrix element in question depends on the longitudinal momentum transfer
$\Delta^+=(p_{\cal M}-p_N) \cdot n$
between a baryon and a meson characterized by the skewness variable
$
\xi= -\frac{(p_{\cal M}-p_N) \cdot n}{(p_{\cal M}+p_N) \cdot n}
$
and by a transverse momentum transfer
$\vec \Delta_T$.

For the QCD evolution equations obeyed by baryon-to-meson TDAs one distinguishes the
Efremov-Radyushkin-Brodsky-Lepage (ERBL)-like domain in which all three  momentum
fractions of quarks
are positive and two kinds of Dokshitzer-Gribov-Lipatov-Altarelli-Parisi (DGLAP)-like regions
in which either one or two momentum fractions of quarks are negative.

\begin{figure}
 \begin{center}
 \includegraphics[width=0.3\textwidth]{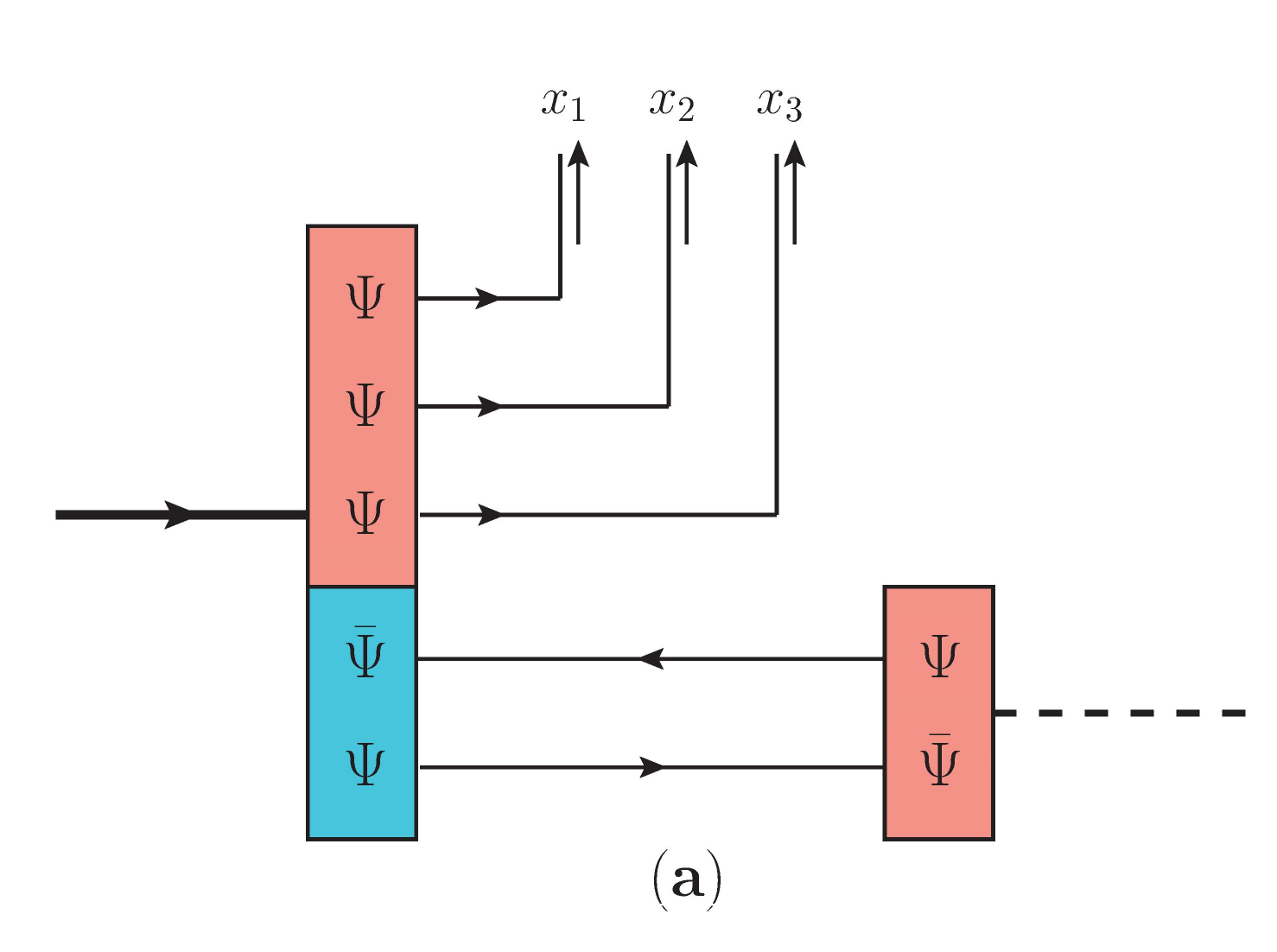} \ \ \
 \includegraphics[width=0.3\textwidth]{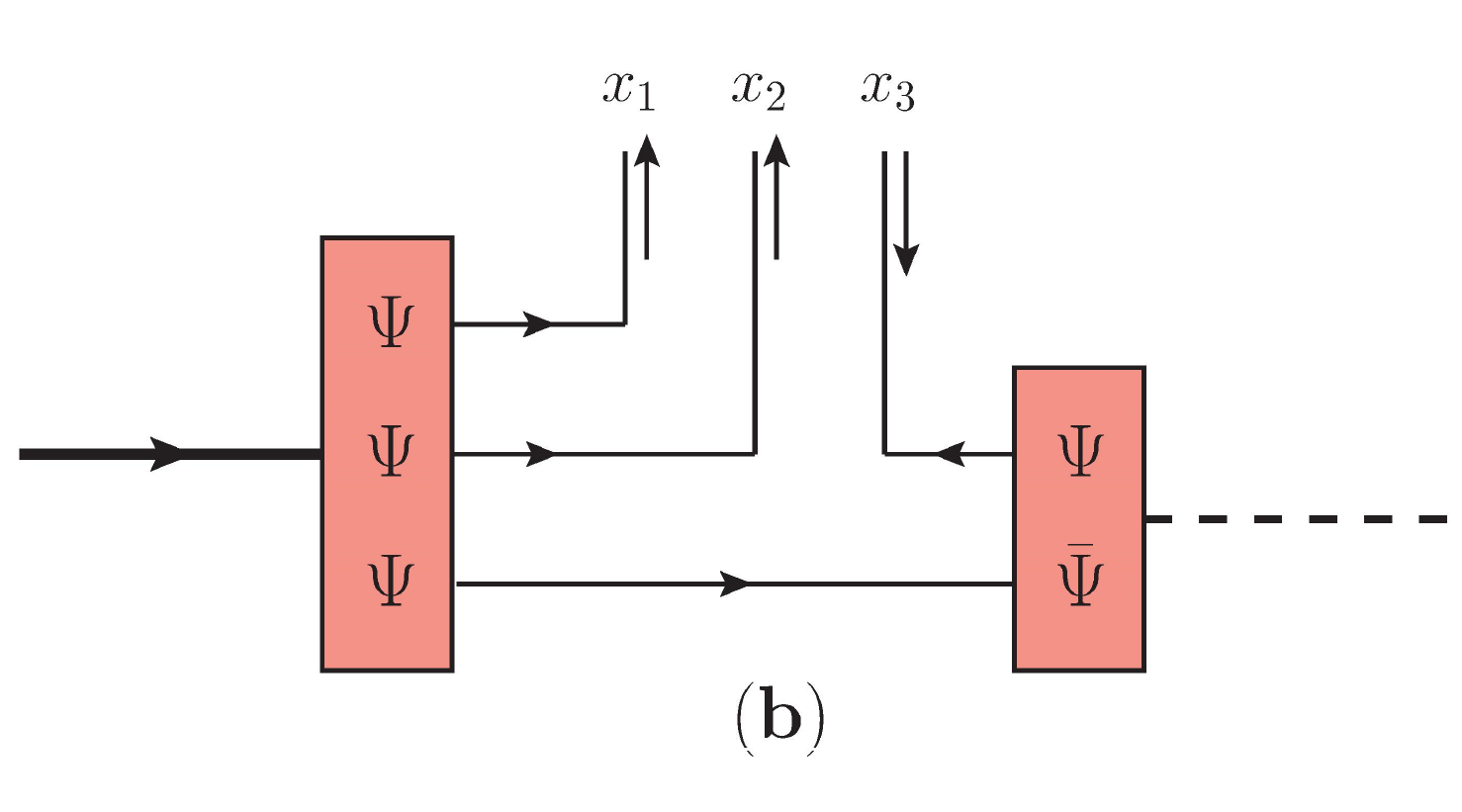} \ \ \
 \includegraphics[width=0.3 \textwidth]{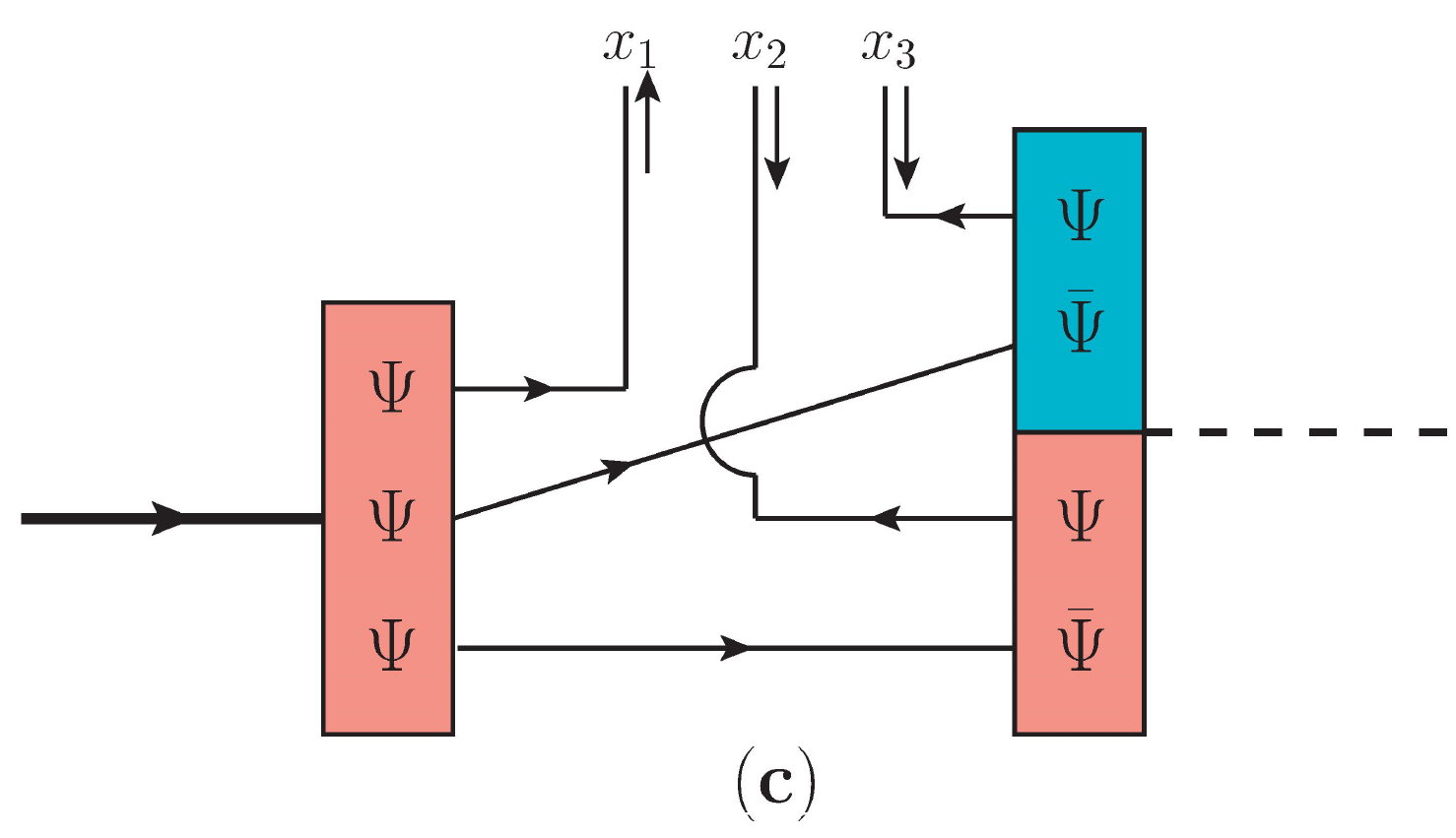}
 \caption{Interpretation of baryon-to-meson TDAs at low normalization scale.
    Small vertical arrows show the flow of the momentum.
     {\bf  (a):} Contribution in the ERBL region (all $x_i$ are positive);
    {\bf  (b):} Contribution in the DGLAP I region (one of $x_i$  is negative).
    {\bf  (c):} Contribution in the DGLAP II region (two  $x_i$  are negative).
     }
 \label{Fig_X}
 \end{center}
 \end{figure}

The physical picture encoded in baryon-to-meson TDAs is conceptually close to that
contained in baryon GPDs and baryon DAs. Baryon-to-meson TDAs  characterize partonic correlations inside a baryon and give access to the momentum distribution
of the baryonic number inside a baryon. The same operator also defines the nucleon DA,
which can be seen as a limiting case of baryon-to-meson TDAs with the meson state replaced by the vacuum.
In the language of the Fock state decomposition, baryon-to-meson TDAs are not restricted to the lowest Fock state
as DAs. They rather probe the non-minimal Fock components with additional
quark-antiquark pair:
\begin{eqnarray}
&&
| {\rm Nucleon} \rangle= |\Psi \Psi \Psi \rangle+ |\Psi \Psi \Psi; \,  \bar{\Psi} \Psi \rangle+....\;; ~~~~
| {\rm Meson} \rangle= |\bar{\Psi}\Psi \rangle+ |\bar{\Psi}\Psi; \, \bar{\Psi} \Psi \rangle+....\;
\end{eqnarray}
depending on the particular support region in question
(see Fig.~\ref{Fig_X}). Note that this interpretation can be
justified only at a very low normalization scale and can be
significantly altered at higher scales due to the evolution effects.

Similarly to GPDs, by Fourier transforming baryon-to-meson TDAs to the
impact parameter space
($\vec \Delta_T \to \vec b_T$),
one obtains additional insight on the baryon structure in the transverse plane.
This allows one to perform the femto-photography of hadrons
\cite{Ralston:2001xs}
from a new perspective.

\section{Parametrization and phenomenological models for baryon-to-meson TDAs}
\label{sec:1}

Although baryon-to-meson TDAs can be introduced for all types of baryons and mesons, we
start our consideration from the simplest case of nucleon-to-pion 
($\pi N$) 
TDAs.
For a given flavor contents ({\it e.g.} 
$uud$ 
proton-to-$\pi^0$ 
TDA) the parametrization
of the leading twist-$3$
$\pi N$
TDA involves
$8$
invariant functions,
each depending on the
$3$
longitudinal momentum fractions
$x_i$,
the skewness parameter
$\xi$,
momentum transfer squared
$\Delta^2$
as well as on the factorization scale
$\mu^2$:
\bea
&&
4 (p \cdot n)^3 \int \left[ \prod_{j=1}^3 \frac{d \lambda_j}{2 \pi}\right]
e^{i \sum_{k=1}^3 x_k \lambda_k (p \cdot n)}
 \langle     \pi^0(p_\pi)|\,  \varepsilon_{c_1 c_2 c_3} u^{c_1}_{\rho}(\lambda_1 n)
u^{c_2}_{\tau}(\lambda_2 n)d^{c_3}_{\chi}(\lambda_3 n)
\,|N^p(p_N,s_N) \rangle
\nonumber \\ &&
= \delta(x_1+x_2+x_3-2 \xi) i \frac{f_N}{f_\pi}\Big[  V^{(p\pi^0)}_{1}(x_{1,2,3}, \xi,\Delta^2)  (  \hat{p} C)_{\rho \tau}(U^+)_{\chi}
\nonumber \\ &&
+A^{(p\pi^0)}_{1}(x_{1,2,3}, \xi,\Delta^2)  (  \hat{p} \gamma^5 C)_{\rho \tau}(\gamma^5 U^+ )_{\chi}
%\nonumber \\ &&
 +T^{(p\pi^0)}_{1}(x_{1,2,3}, \xi,\Delta^2)  (\sigma_{p\mu} C)_{\rho \tau }(\gamma^\mu U^+ )_{\chi}
 \nonumber \\ &&
 + m_N^{-1} V^{(p\pi^0)}_{2} (x_{1,2,3}, \xi,\Delta^2)
 ( \hat{ p}  C)_{\rho \tau}( \hat{\Delta}_T U^+)_{\chi}
+ m_N^{-1}
 A^{(p\pi^0)}_{2}(x_{1,2,3}, \xi,\Delta^2)  ( \hat{ p}  \gamma^5 C)_{\rho\tau}(\gamma^5  \hat{\Delta}_T  U^+)_{\chi}
  \nonumber \\ &&
+ m_N^{-1} T^{(p\pi^0)}_{2} (x_{1,2,3}, \xi,\Delta^2) ( \sigma_{p\Delta_T} C)_{\rho \tau} (U^+)_{\chi}
%\nonumber \\
+  m_N ^{-1}T^{(p\pi^0)}_{3} (x_{1,2,3}, \xi,\Delta^2) ( \sigma_{p\mu} C)_{\rho \tau} (\sigma^{\mu\Delta_T}
 U^+)_{\chi}
 \nonumber \\ &&
+ m_N^{-2} T^{(p\pi^0)}_{4} (x_{1,2,3}, \xi,\Delta^2)  (\sigma_{p \Delta_T} C)_{\rho \tau}
(\hat{ \Delta}_T U^+)_{\chi} \Big].
 \label{Old_param_TDAs}
\eea
Here
$f_\pi=93$ MeV
is the pion weak decay constant and
$f_N$
determines the value of the nucleon wave function at the origin.
Throughout this paper we adopt Dirac's ``hat'' notation
$\hat{v} \equiv v_\mu \gamma^\mu$;
$\sigma^{\mu\nu}= \frac{1}{2} [\gamma^\mu, \gamma^\nu]$; $\sigma^{v \mu} \equiv v_\lambda \sigma^{\lambda \mu}$;
$C$
is the charge conjugation matrix and
$U^+= \hat{p} \hat{n} \, U(p_N,s_N)$
is the large component of the nucleon spinor. The parametrization
(\ref{Old_param_TDAs}), 
originally introduced in
Ref.~\cite{Lansberg:2007ec},
is extremely convenient for the phenomenological applications
since in the limit
$\Delta_T \to 0$
just three invariant amplitudes
$V^{(p \pi^0)}_{1}$,
$A^{(p \pi^0)}_{1}$
and
 $T^{(p\pi^0)}_{1}$
do survive.
An alternative parametrization which is better suited to address
the symmetry properties of
$\pi N$
TDAs was introduced in
\cite{Pire:2011xv}.
In particular, within this latter parametrization
$\pi N$
TDAs
turn to satisfy the polynomiality property of the Mellin moments in
$x_i$
in its most simple form. The relation between the two
parametrizations in presented in the Appendix~C of
\cite{Pire:2011xv}.

Working out the physical normalization of baryon-to-meson TDAs and building up
consistent phenomenological models for them represents a considerable task. Below
we present a short overview of the approaches available at the present moment.

The first attempt to provide a model for
$\pi N$
TDAs relying on chiral dynamics
was performed in
\cite{Lansberg:2007ec}
(see
Ref.~\cite{Pire:2011xv}
for the detailed derivation).
The soft-pion theorem
\cite{Pobylitsa:2001cz,Braun:2006td}
allows to express
$\pi N$ TDAs
$V^{(p \pi)}_{1}$,
$A^{(p \pi)}_{1}$
and
$T^{(p\pi)}_{1}$
in the soft pion limit
($\xi=1$; $\Delta^2=m_N^2$)
in terms of the $3$ leading twist nucleon DAs
$\{V^p, A^p, T^p\}(y_1,y_2,y_3) \equiv \{V^p, A^p, T^p\}(y_i)$. %, $A^p$ and $T^p$.
This results in a simple model for
$\pi N$
TDAs summarized in the Erratum to
\cite{Lansberg:2007ec}:%
\footnote{Note the relative signs and the overall factor
$\frac{1}{2}$
missed in the  Erratum to
\cite{Lansberg:2007ec}.}
\bea
%&&
\{V^{(p \pi^0)}_{1}, \, A^{(p \pi^0)}_{1} \}(x_i,\, \xi)=
-\frac{1}{2} \times \frac{1}{4 \xi}
\{ V^p,\, A^p\} \left( \frac{x_i}{2 \xi} %,
%\frac{x_2}{2 \xi}, \frac{x_3}{2 \xi}
\right);
%\nonumber \\ &&
\ \ \
T^{(p \pi^0)}_{1} (x_i,\, \xi)=
-\frac{1}{2} \times \frac{3}{4 \xi}
T^p \left( \frac{x_i}{2 \xi}
%, \frac{x_2}{2 \xi}, \frac{x_3}{2 \xi}
\right);
\label{SoftPionTDAmodel_revised}
\eea
Despite its obvious drawbacks
(like the very narrow validity range limited to the
close vicinity of the threshold and lack of an intrinsic $\Delta^2$-dependence)
this simple model
for the first time provided a quantitative estimate
of the physical normalization for
$\pi N$
TDAs.
In particular, the predictions of the revised soft-pion-limit model
(\ref{SoftPionTDAmodel_revised})
were recently employed in the first feasibility studies
\cite{Singh:2014pfv}
for
accessing
$\pi N$
TDAs with \={P}ANDA through
$\bar{p}p \to \gamma^* \pi^0$.

\begin{figure}
 \begin{center}
 \includegraphics[width=0.2\textwidth]{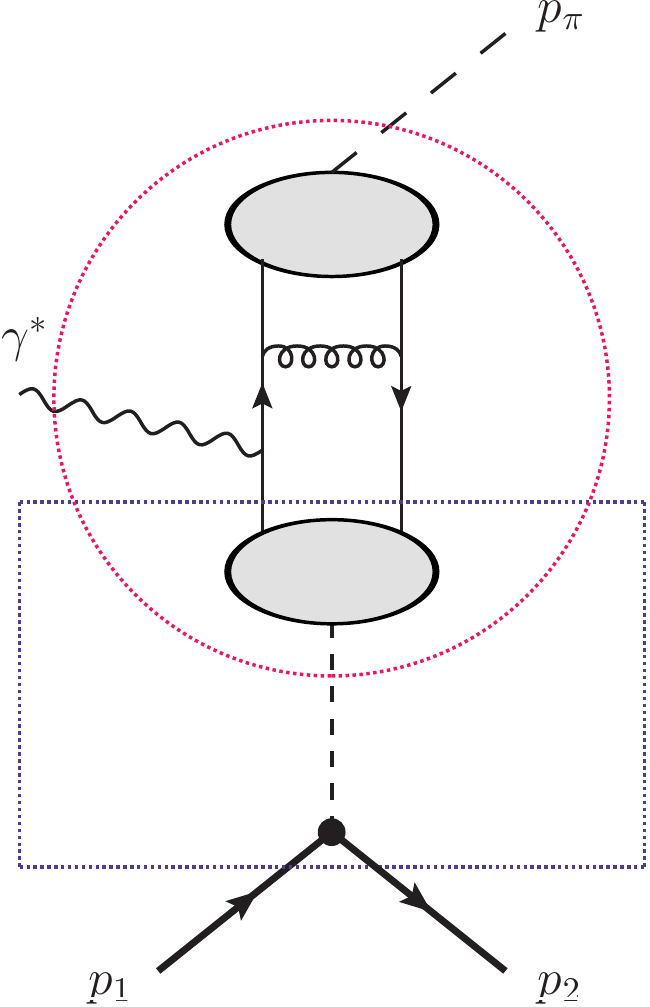} \ \ \ \ \ \ \ \ \ ~~~~~~~~~~~~~~~
  \includegraphics[width=0.2\textwidth]{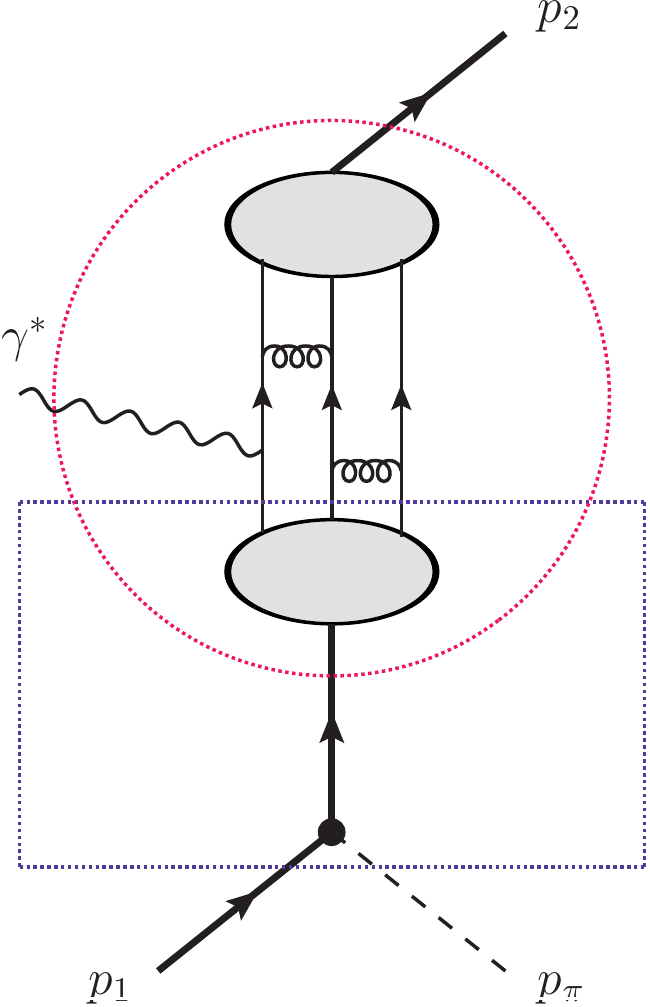}
      \caption{{\bf Left:} pion pole exchange model for the polarised GPD $\tilde{E}$; lower and upper blobs depict pion DAs;  the dashed circle contains
     a typical LO graph for the pion electromagnetic form factor in perturbative QCD; the rectangle contains the pion pole contribution into GPD.
    {\bf Right:} nucleon pole exchange model for $\pi N$ TDAs;  dashed circle contains a typical LO graph for the nucleon electromagnetic form factor in perturbative QCD; the rectangle contains the nucleon pole contribution into $\pi N$ TDAs.  }
\label{Fig_Nucleon_pole}
\end{center}
\end{figure}

Another simple model for 
$\pi N$ 
TDAs suggested in Ref.~\cite{Pire:2011xv}
accounts for the contribution of 
the cross-channel
nucleon exchange. As one can see from Fig.~\ref{Fig_Nucleon_pole},
this model is conceptually similar to the
pion exchange model for the polarized nucleon GPD
$\tilde{E}$
suggested in Ref.~\cite{Goeke:2001tz}.
With the use of the
$\pi N$
TDA parametrization
(\ref{Old_param_TDAs})
the nucleon pole model reads:
\bea
&&
\big\{ V_1, \, A_1 , \, T_1  \big\}^{(p \pi^0)} ( %x_1,x_2,x_3,
x_i, \xi,\Delta^2)\Big|_{N(940)}
%\nonumber \\ &&
 =\Theta_{\rm ERBL}(x_k)
 %,x_2,x_3)
 \times  \frac{g_{\pi NN} \,m_N f_\pi}{\Delta^2-m_N^2}    \frac{1}{(2 \xi) } \frac{1-\xi}{1+\xi}
 \big\{ V^p,\,A^p, \,T^p  \big\}\left( \frac{x_i}{2 \xi}
 %, \frac{x_2}{2 \xi}, \frac{x_3}{2 \xi}
 \right);
 \nonumber \\  &&
 \big\{ V_2, \, A_2 , \, T_2, \, T_3  \big\}^{(p \pi^0)} ( %x_1,x_2,x_3,
x_i, \xi,\Delta^2)\Big|_{N(940)}
%\nonumber \\ &&
=\Theta_{\rm ERBL}(x_k) \times    \frac{g_{\pi NN} \, m_N f_\pi}{\Delta^2-m_N^2}    \frac{1}{(2 \xi) }
 \big\{ V^p,\,A^p, \,T^p, \, T^p  \big\}\left( \frac{x_i}{2 \xi}
 %, \frac{x_2}{2 \xi}, \frac{x_3}{2 \xi}
 \right);
\nonumber \\  &&
   T_4^{(p \pi^0 )} ( %x_1,x_2,x_3,
x_i, \xi,\Delta^2)\Big|_{N(940)}=0;
\nonumber \\  &&
\big\{ V_{1,2}, \, A_{1,2} , \, T_{1,2,3,4}  \big\}^{(p \pi^+)} ( %x_1,x_2,x_3,
x_i, \xi,\Delta^2)\Big|_{N(940)}= -\sqrt{2} \big\{V_{1,2}, \, A_{1,2} , \, T_{1,2,3,4}    \big\}^{(p \pi^0 )} ( %x_1,x_2,x_3,
x_i, \xi,\Delta^2)\Big|_{N(940)}.
\label{Nucleon_exchange_contr_VAT}
\eea
Here
$
\Theta_{\rm ERBL}(x_k)  \equiv  \prod_{k=1}^3 \theta(0 \le x_k \le 2 \xi)
$
ensures the pure ERBL-like support of TDAs and
$g_{\pi NN}\approx 13$
is the pion-nucleon phenomenological coupling.
This turns to be a consistent model for
$\pi N$
TDAs in the ERBL-like region and satisfies the polynomiality conditions 
and the appropriate symmetry relations.

By an obvious change of couplings the model
(\ref{Nucleon_exchange_contr_VAT})
can be generalized to the
case of other light mesons
($\eta$, $\eta'$, $K$, ... {\it etc.}).
Also it is not necessarily limited to the contribution
of the cross-channel nucleon exchange.
In Ref.~\cite{Pire:2011xv}
the contribution of the cross-channel
$\Delta(1232)$
into
$\pi N$
TDAs was worked out explicitly. Finally, very recently in
Ref.~\cite{Pire:2015kxa}
the cross-channel nucleon exchange model was generalized for the case of
nucleon-to-vector meson TDAs.

Still the aforementioned baryon-to-meson TDA model describes TDAs
only within the ERBL-like support region. To get a model defined
on the complete support domain one may rely on the spectral representation
for baryon-to-meson TDAs in terms of quadruple distributions suggested in Ref.~\cite{Pire:2010if}:
\begin{eqnarray}
\label{Spectral_for_GPDs_x123}
&&
H^{({\cal M} N)}(x_1,\,x_2,\,x_3=2 \xi -x_1-x_2,\,\xi,\,t)  \\ && =
\left[
\prod_{i=1}^3
\int_{\Omega_i} d \beta_i d \alpha_i
\right]
%\int_{\Omega_2} d \beta_2 d \alpha_2
%\int_{\Omega_3} d \beta_3 d \alpha_3
\delta(x_1-\xi-\beta_1-\alpha_1 \xi) \,
\delta(x_2-\xi-\beta_2-\alpha_2 \xi) \,
%\delta(x_3-\xi-\beta_3-\alpha_3 \xi) \,
%\nonumber \\ &&
%\times
\delta(\sum_i \beta_i)
\delta(\sum_i \alpha_i+1)
 f(\beta_i, \, \alpha_i,\,t), \nonumber
%\nonumber \\ &&
\end{eqnarray}
where
$\Omega_{i}$
denote three copies of the usual domain in the spectral parameter space.
The spectral density
$f$
is an arbitrary function of six variables, which are  subject to two constraints 
and therefore effectively is a quadruple distribution.

Similarly to the familiar double distribution representation for GPDs  the
quadruple distribution representation
(\ref{Spectral_for_GPDs_x123})
for TDAs turns to be the most
general way to implement the support properties of TDAs as well as the
polynomiality property for the $x_i$-Mellin moments,
which is a direct consequence of Lorentz invariance
(see Ref.~\cite{Lansberg:2011aa}).

Contrarily to GPDs, TDAs do not possess the comprehensive forward limit
$\xi \to 0$.
This complicates the construction of a phenomenological Ansatz for quadruple
distributions.
However, a partial solution was proposed for the case of
$\pi N$ 
TDAs. In this case one can rely on the complementary
$\xi \to 1$ limit,
in which, as  already discussed above,
$\pi N$
TDAs
$V_1^{\pi N}$,
$A_1^{\pi N}$,
$T_1^{\pi N}$
are constrained by chiral dynamics
(see eq.~(\ref{SoftPionTDAmodel_revised}) for
$\xi=1$).
In Ref.~\cite{Lansberg:2011aa}
we proposed a factorized Ansatz designed as a universal profile
function multiplying the nucleon DA combination to which the 
$\pi N$ 
TDA in question reduces in the
$\xi=1$
limit.

In loose words our Ansatz for quadruple distributions
is based on ``skewing'' the
$\xi=1$
limit.
At this point we are similar to the famous Radyushkin
factorized Ansatz for double distributions
\cite{Musatov:1999xp}.
In that case rather the
$\xi=0$ 
limit, where GPDs reduce to usual
parton densities, is ``skewed'' to provide a non trivial $\xi$-dependence for
GPDs. For  practical details we refer the reader to
Ref.~\cite{Lansberg:2011aa}.

Also similarly to the GPD case, in order to satisfy the polynomiality
condition in its complete form, the spectral part
(\ref{Spectral_for_GPDs_x123})
is to be complemented by a $D$-term-like contribution defined solely in 
the ERBL-like region and responsible for the highest possible power of
$\xi$ 
occurring for a given
$x_i$-Mellin moment.
The simplest possible model for such a $D$-term is
the contribution of the
% $u$-channel 
nucleon exchange
into 
$\pi N$ 
TDAs
(\ref{Nucleon_exchange_contr_VAT}).
Note that we avoid double counting since the
%$u$-channel 
nucleon exchange contribution into
$\pi N$ 
TDAs
$V_1^{(\pi N)}$,
$A_1^{(\pi N)}$,
$T_1^{(\pi N)}$
dies out in the 
$\xi \to 1$ 
limit.

In this way we come to the so-called ``two component'' model
for $\pi N$ TDAs
\cite{Lansberg:2011aa}
which contains a spectral part fixed from the
$\xi \to 1$ limit and a $D$-term-like contribution coming from
 the cross-channel 
nucleon exchange. This model provides a lively
$x_i$ and $\xi$ dependence for $\pi N$ TDAs. However, integrating  a flexible but reasonable $t-$dependence still remains an open question.

Ref.~\cite{Lansberg:2011aa} contains the first attempts of phenomenological application
of the ``two component'' 
$\pi N$ 
TDA model for the description of backward pion electroproduction off nucleons for the typical JLab kinematical conditions.
$\pi N$ TDAs within this model do not nullify at the cross-over trajectories separating
the ERBL-like and DGLAP-like TDA support regions. This results in the non-zero contribution
into the imaginary part of the relevant elementary amplitude.
The observable quantity sensitive to this issue is the transverse target single spin
asymmetry (STSA) \cite{Lansberg:2010mf}.
The ``two component'' 
$\pi N$ 
TDA model predicts a sizeable value of STSA for
$x_B$ 
typical for JLab$@ 12$ GeV conditions.

\section{Phenomenology}
%Lack of place forbids us to detail our works on the phenomenology of TDAs in various %reactions. We thus refer to Refs. \cite{Lansberg:2007ec, Lansberg:2007se, Lansberg:2012ha, %Lansberg:2011aa, PsiTDA} for detailed results.

\subsection{Baryon-to-meson TDAs from backward meson electroproduction at JLab}

\begin{figure}
 \begin{center}
 \includegraphics[width=0.35\textwidth]{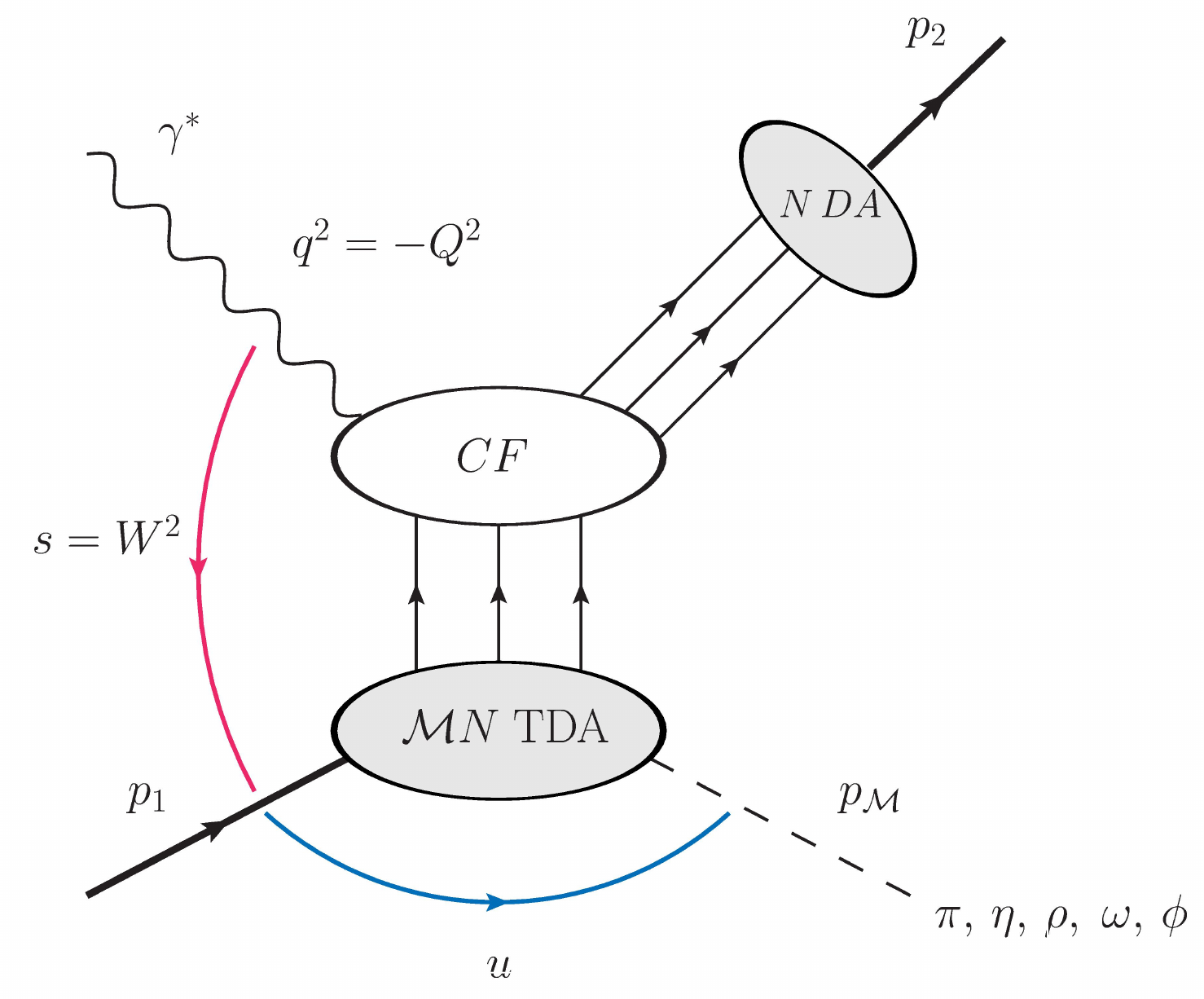}
  \includegraphics[width=0.5\textwidth]{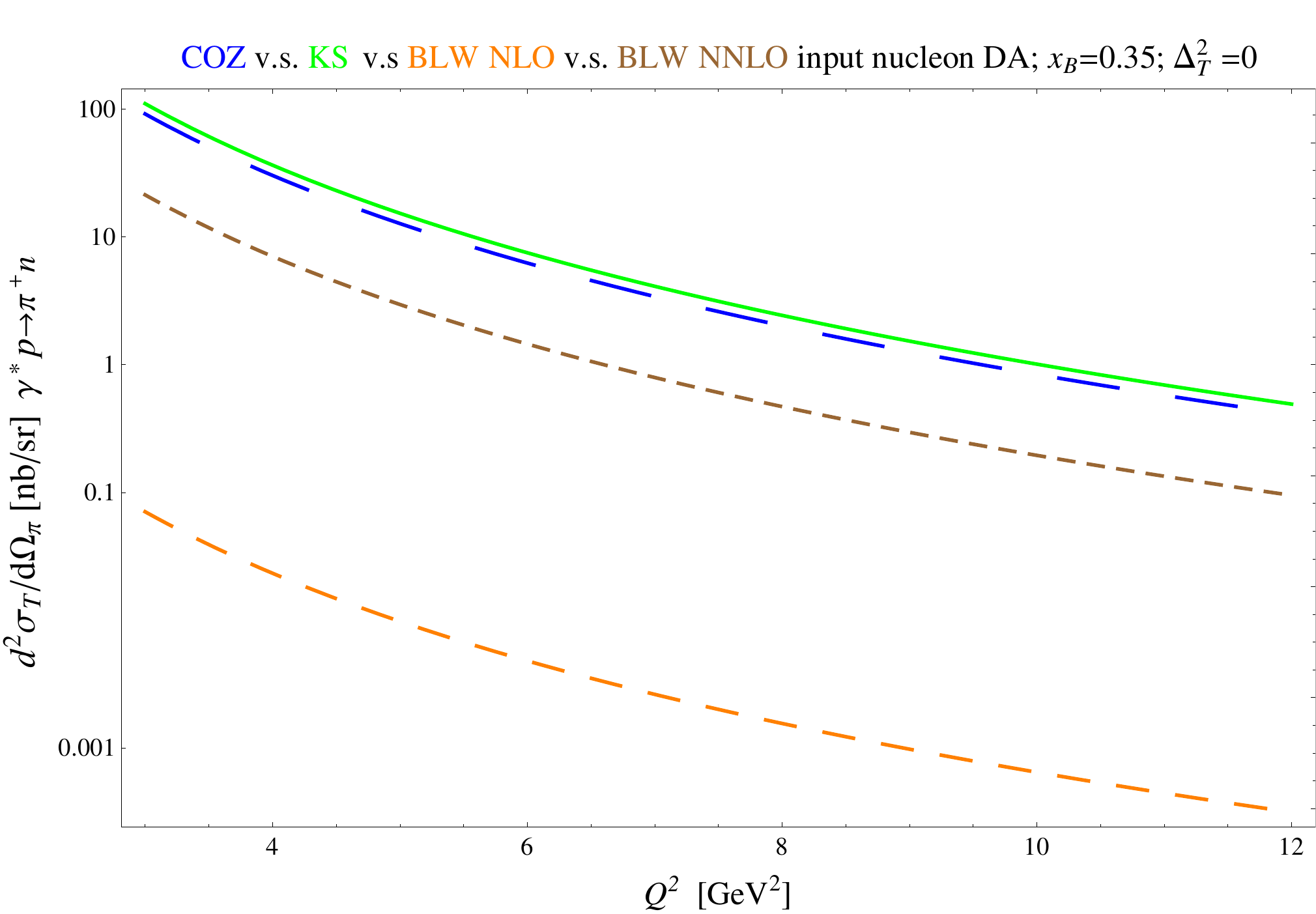}
 \caption{{\bf Left}: Collinear factorization  of $\gamma^* N \to N {\cal M}$  in the  near-backward  kinematics regime.  ${\cal M}N$ TDA stands for the nucleon-to-meson TDA; $N$ DA stands for the nucleon DA; $CF$  denotes hard subprocess amplitudes (coefficient functions).
    {\bf Right:}   Unpolarized cross section $\frac{d^2 \sigma_T}{d \Omega_\pi}$ (in nb/sr)
    for strictly backward $\gamma^* p \to \pi^+ n$ as a function of $Q^2$
for
$x_B=0.35$
with various input phenomenological nucleon DA solutions: COZ (long blue dashes);
KS (solid green line);   BLW NLO
(medium orange dashes) and NNLO modification~\cite{Lenz:2009ar} of BLW (short brown dashes).}
\label{Fig_fact_Jlab}
  \end{center}
\end{figure}

Within the generalized Bjorken limit, in which
$Q^2=-q^2$ and $s=(q+p_1)^2=W^2$ are large,  $x_{B} \equiv \frac{Q^2}{2 p_1 \cdot q}$ is fixed and
the $u$-channel momentum transfer squared is small compared to $Q^2$ and $s$
($|u| \equiv |(p_2-p_1)^2| \ll Q^2,\,s$)
the amplitude of the hard subprocess of the exclusive electroproduction of mesons off nucleons \be
e(k)+N(p_1,s_1) \to
\left(\gamma^*(q, \lambda_\gamma)+N(p_1,s_1) \right) + e(k') \to e(k')+N(p_2,s_2)+ {\cal M}(p_{\cal M}),
\label{hard_meson_production}
\ee
is supposed to admit a collinear factorized description in terms of
nucleon-to-meson TDAs and nucleon DAs, as it is shown on Fig.~\ref{Fig_fact_Jlab}.
The small 
$u$ 
corresponds to the meson being produced in the near-backward direction in the
$\gamma^* N$ 
center-of-mass system (CMS). This regime, referred as the near-backward kinematics, is complementary to the more conventional near-forward kinematical regime 
($Q^2=-q^2$, $s$ -- large, 
$x_{B}$ -- fixed, 
$|t| \equiv |(p_2-p_1)^2| \ll Q^2, \,s $)
in which the factorized description in terms of GPDs and meson DAs applies to hard meson production subprocess.

The details of the formalism for the case of backward electroproduction of light pseudoscalar mesons
($\pi$, $\eta$)
and rough cross section estimates for the kinematics conditions of JLab can be found in Refs.~\cite{Lansberg:2007ec, Lansberg:2011aa}.
A generalization for the case of vector mesons
($\rho$, $\omega$, $\phi$)
was proposed in
\cite{Pire:2015kxa}.

As an example of our predictions, on the right panel of Fig.~\ref{Fig_fact_Jlab} 
we present out estimates of the backward
$\gamma^* p \to \pi^+ n$
cross section within the factorized description involving
$\pi N$
TDAs using the cross channel nucleon exchange model
(\ref{Nucleon_exchange_contr_VAT}).
The cross section turns out to very sensitive to the form of the input
phenomenological solution for nucleon DA. We show the results
for Chernyak-Ogloblin-Zhitnitsky (COZ)
\cite{Chernyak:1987nv}
(long blue dashes);
King-Sachrajda (KS)
\cite{King:1986wi}
(solid green line);
Braun-Lenz-Wittmann next-to-leading-order (BLW NLO)
\cite{Braun:2006hz}
(medium orange dashes) and
NNLO modification~\cite{Lenz:2009ar}
of BLW (short brown dashes).
The solutions close to the asymptotic form of the nucleon DA
$\phi(y_i)=120 y_1 y_2 y_3$
result in a very small cross section, while those significantly different from the
asymptotic form like COZ and KS result in larger cross sections. 
%from the suggested perturbative QCD mechanism. 
We refer the reader {\it e.g.} to the
discussion in
Ref.~\cite{Stefanis_DrNauk}
on various phenomenological inputs for nucleon DAs.

A first experimental signal may have been detected at JLab
\cite{Kubarovskiy:2012yz}
in backward kinematics for
$e^-\,N \to e^- \pi  \,N$
and for
$e^-\,N \to e^- \omega  \,N$
\cite{HuberPrivate}.
We expect the accessible kinematical domain to be more adequate to a factorized 
leading twist analysis in higher energy experiments at 
JLab@12 GeV.

\subsection{Study of baryon-to-meson TDAs at \={P}ANDA}

\begin{figure}
 \begin{center}
 \includegraphics[width=0.35\textwidth]{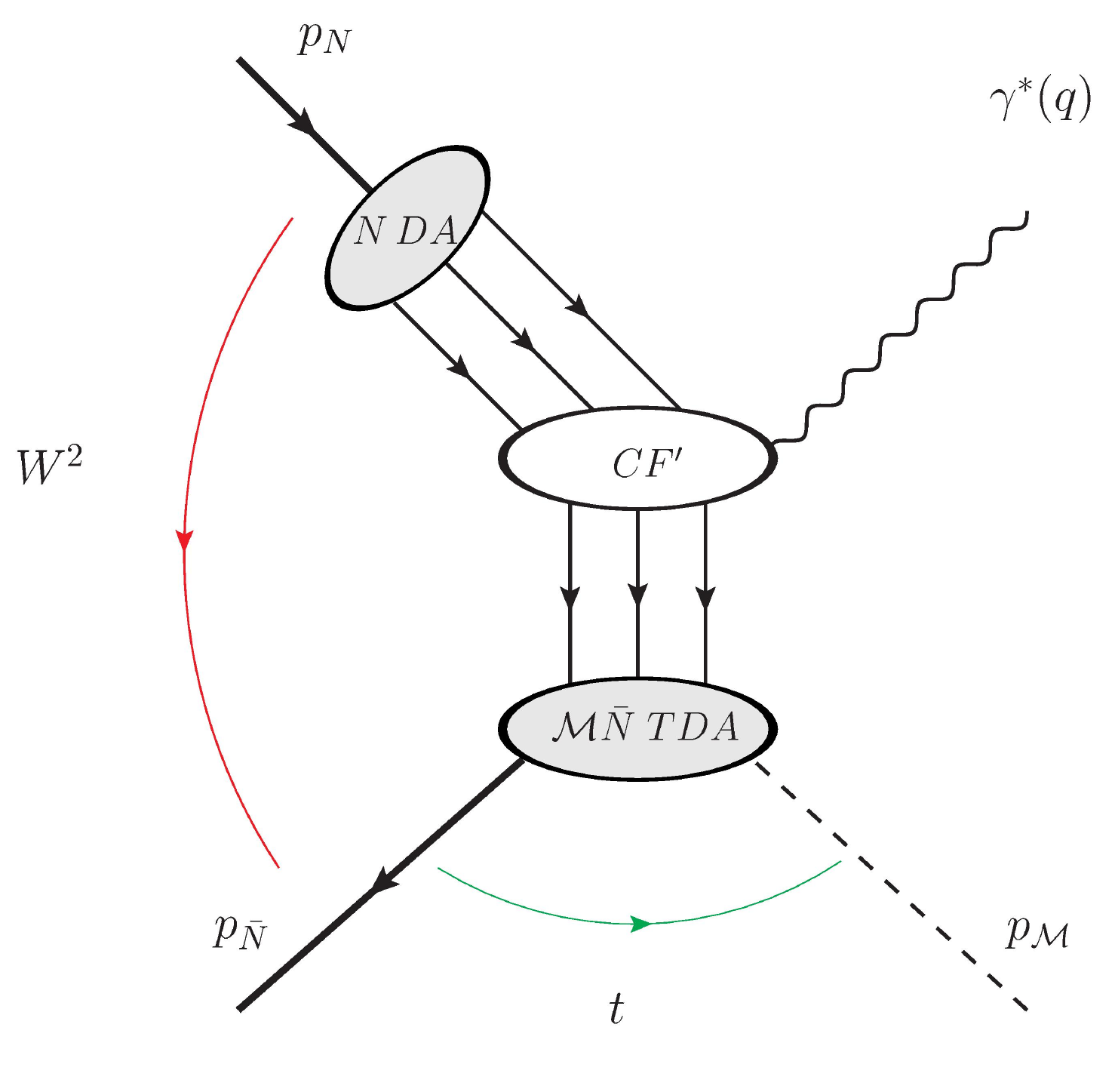} \ \ \  ~~~~~~~~~~~~~~~
  \includegraphics[width=0.4\textwidth]{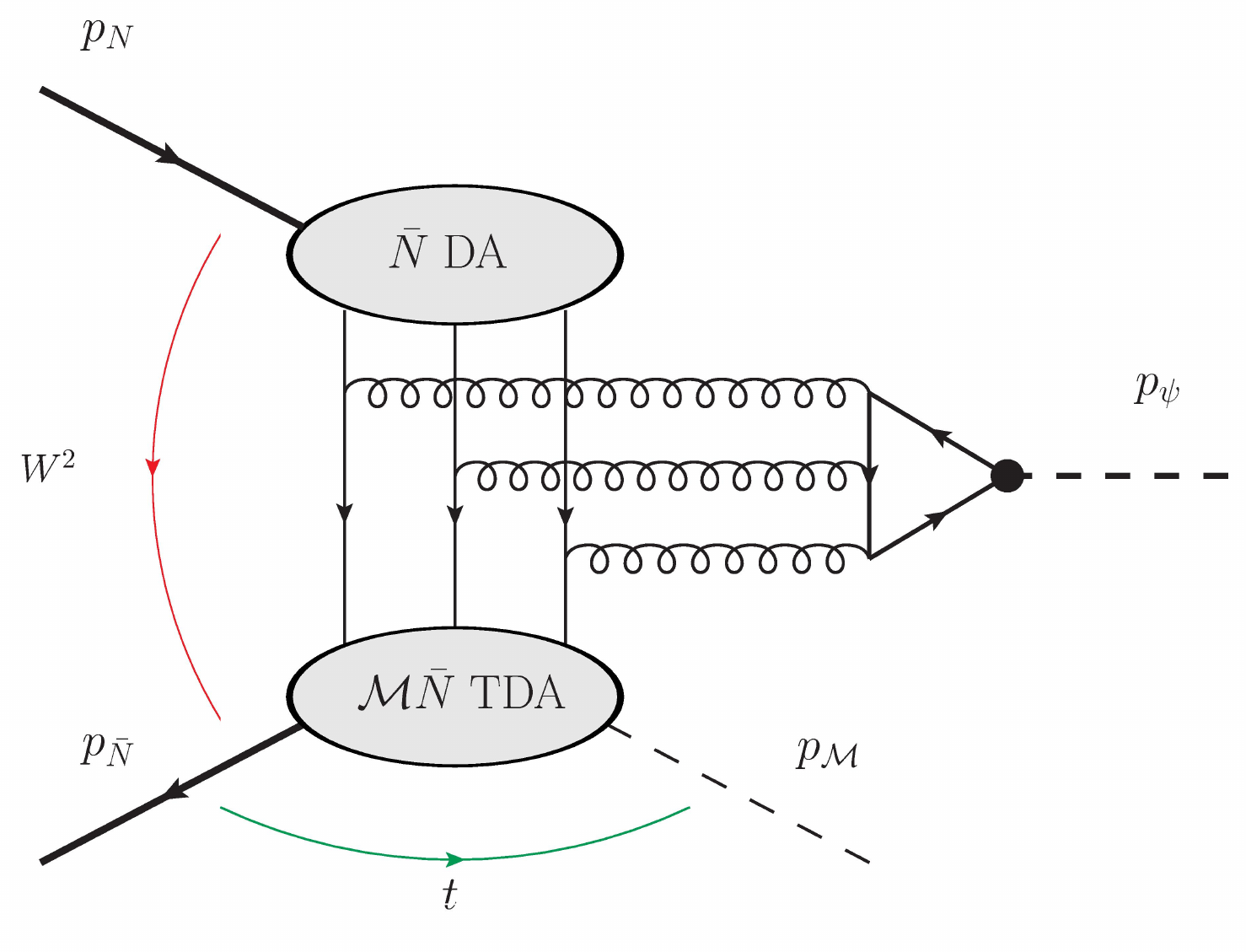}
     \caption{{\bf Left:}
      Collinear factorization  of the annihilation process
$N \bar{N} \to \gamma^*(q) \mathcal{M}(p_\mathcal{M})$.
(near-forward  kinematics).
{\bf Right:}  Collinear factorization  of the annihilation
     process $N + \bar{N} \to J/\psi + {\cal M}$  (near-forward  kinematics).
     DA stands for the distribution amplitude
of a nucleon;
$\mathcal{M} \bar{N}$
TDA stands for the TDA from an antinucleon to a meson. By thick black dot we denote the light-cone wave function of heavy quarkonium.}
\label{Fig_TDAS_at_PANDA}
\end{center}
\end{figure}

\begin{figure}
\begin{center}
\includegraphics[width=0.4\textwidth]{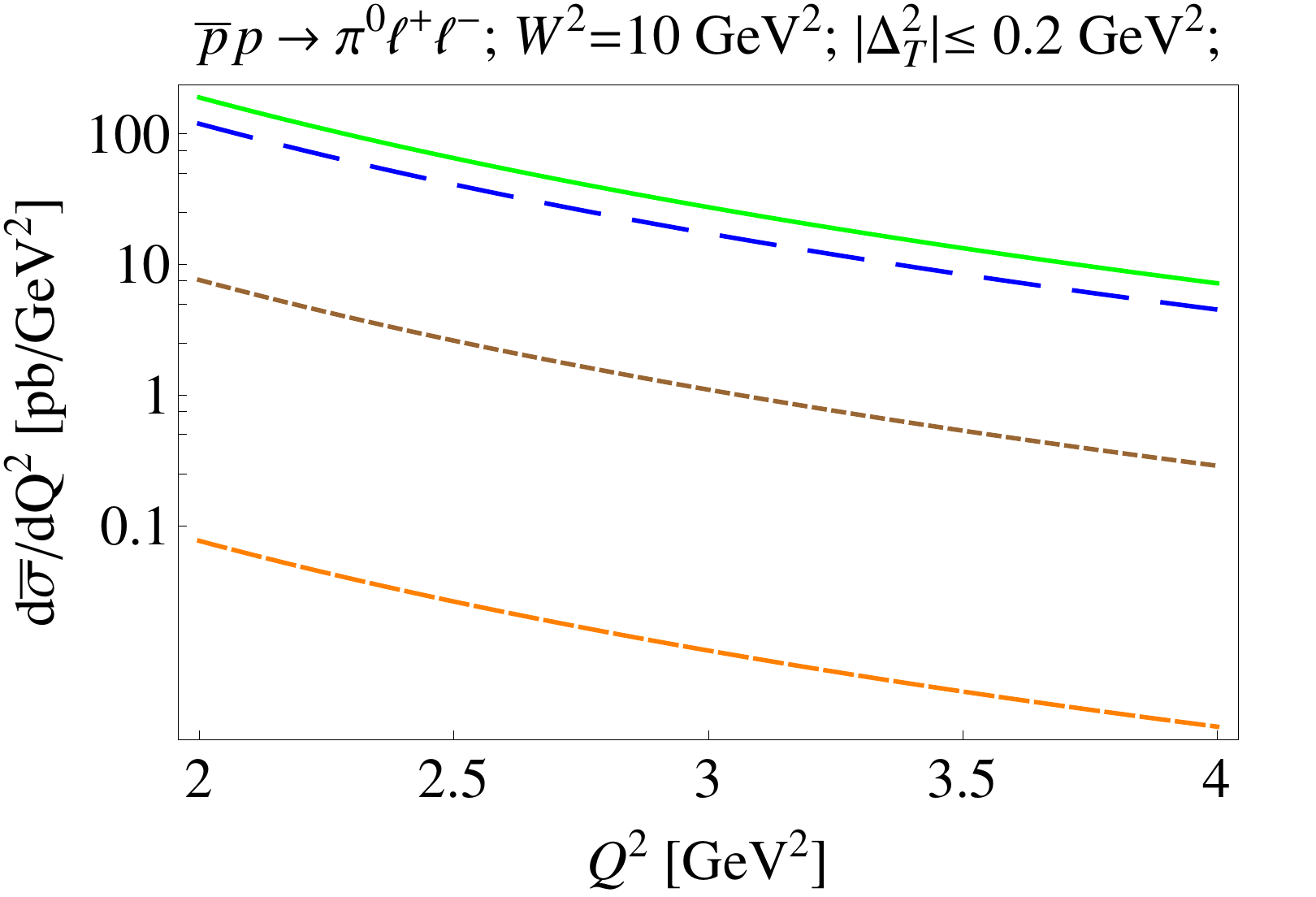} \ \ \ \
\includegraphics[width=0.4\textwidth]{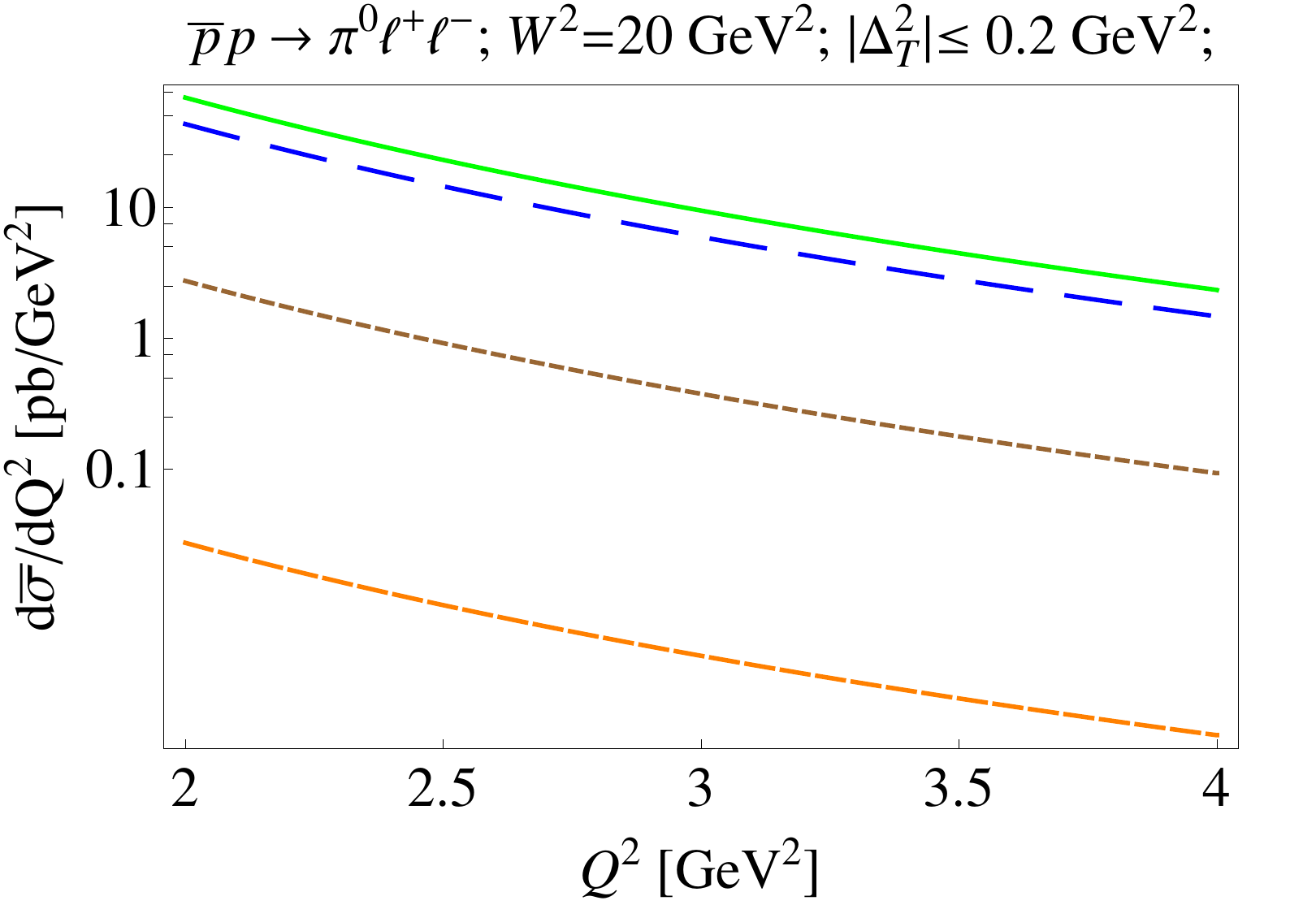}
\caption{ Integrated cross section
$d \bar{\sigma}  /dQ^2$
for
$\bar{p}p \rightarrow \ell^+\ell^- \pi^0$
as a function of
$Q^2$
for
$W^2=10$ GeV$^2$ and $W^2=20$ GeV$^2$
for various phenomenological nucleon DA solutions: COZ (long blue dashes);
KS (solid green line);   BLW NLO
(medium orange dashes) and NNLO modification~\cite{Lenz:2009ar} of BLW (short brown dashes).}
\label{Fig_Dilepton_BkwFwd_CS}       % Give a unique label
\end{center}
\end{figure}

Another tempting possibility to get experimental access to
baryon-to-meson TDAs is to consider the time-like counterpart
of  reaction (\ref{hard_meson_production}):
the nucleon-antinucleon  annihilation into a high invariant mass lepton pair
in association with a light meson
${\mathcal{M}}$:
\begin{equation}
\bar{N} (p_{\bar{N}},s_{\bar{N}})+  N (p_N,s_N) \rightarrow \gamma^*(q)+ {\mathcal{M}}(p_{\mathcal{M}}) \rightarrow \ell^+(p_{\ell^+}) +
\ell^-(p_{\ell^-}) + {\mathcal{M}}(p_{\mathcal{M}}).
\label{BarNNannihilation reaction}
\end{equation}
The factorization mechanism for
(\ref{BarNNannihilation reaction})
suggested in
\cite{Lansberg:2007se,Lansberg:2012ha}
applies within two kinematic regimes referred as the near-forward%
\footnote{With respect to the positive direction defined along the antiproton beam.}
 with
($s=(p_N+ p_{\bar{N}})^2 \equiv W^2$, $Q^2$
large with
$\xi^t=- \frac{(p_{\mathcal{M}}-p_{\bar{N}}) \cdot n^t}{(p_{\mathcal{M}}+p_{\bar{N}}) \cdot n^t}$
fixed; and
$|t|=|(p_{\mathcal{M}}-p_{\bar{N}})^2| \sim 0$)
(see the left panel of Fig.~\ref{Fig_TDAS_at_PANDA})
and the near-backward one, which differs by the obvious change
of kinematical variables
($p_N \to p_{\bar{N}}$, $p_{\bar{N}} \to p_N$, $t \to u$).
The charge conjugation invariance results in perfect symmetry between
the two kinematical regimes.
The suggested reaction mechanism manifests itself through the characteristic forward
and backward peaks of the
$N \bar{N} \to \gamma^* \mathcal{M}$
cross section.
The characteristic features of the TDA-based description of
(\ref{BarNNannihilation reaction})
are the scaling behavior in $1/q^2$ of the cross-section
and the specific  behavior in
$\cos \theta_{\ell}^*$
(by $\theta_{\ell}^*$ 
we denote the lepton polar angle defined in the CMS of the lepton pair)
resulting from the leading twist dominance of the transverse polarization 
of the virtual time-like photon. A detailed feasibility study for accessing
$\pi N$
TDAs through
$\bar p\,p \to  \gamma^* \,\pi^0 \to  e^+ e^- \pi^0$
at \={P}ANDA was performed in
\cite{Singh:2014pfv}.
As a phenomenological input, the predictions of the simple
$\pi N$
TDA model
(\ref{SoftPionTDAmodel_revised})
were employed. The results of this analysis
are promising concerning the experimental perspectives for accessing
$\pi N$
TDAs
within \={P}ANDA@GSI-FAIR experiment.
On Fig.~\ref{Fig_Dilepton_BkwFwd_CS} we present the integrated cross section
$d \bar{\sigma}  /dQ^2$ 
with
$|\Delta_T^2| \le 0.2$~GeV$^2$
for
$\bar{p}p \rightarrow e^+e^- \pi^0$
as a function of
$Q^2$ for $W^2=10$~GeV$^2$ and $W^2=20$~GeV$^2$
within the collinear factorization approach.
We employ the cross channel nucleon exchange model
(\ref{Nucleon_exchange_contr_VAT})
for
$\pi N$
TDAs with various phenomenological  DAs used as inputs.

\begin{figure}
 \begin{center}
 \includegraphics[width=0.45\textwidth]{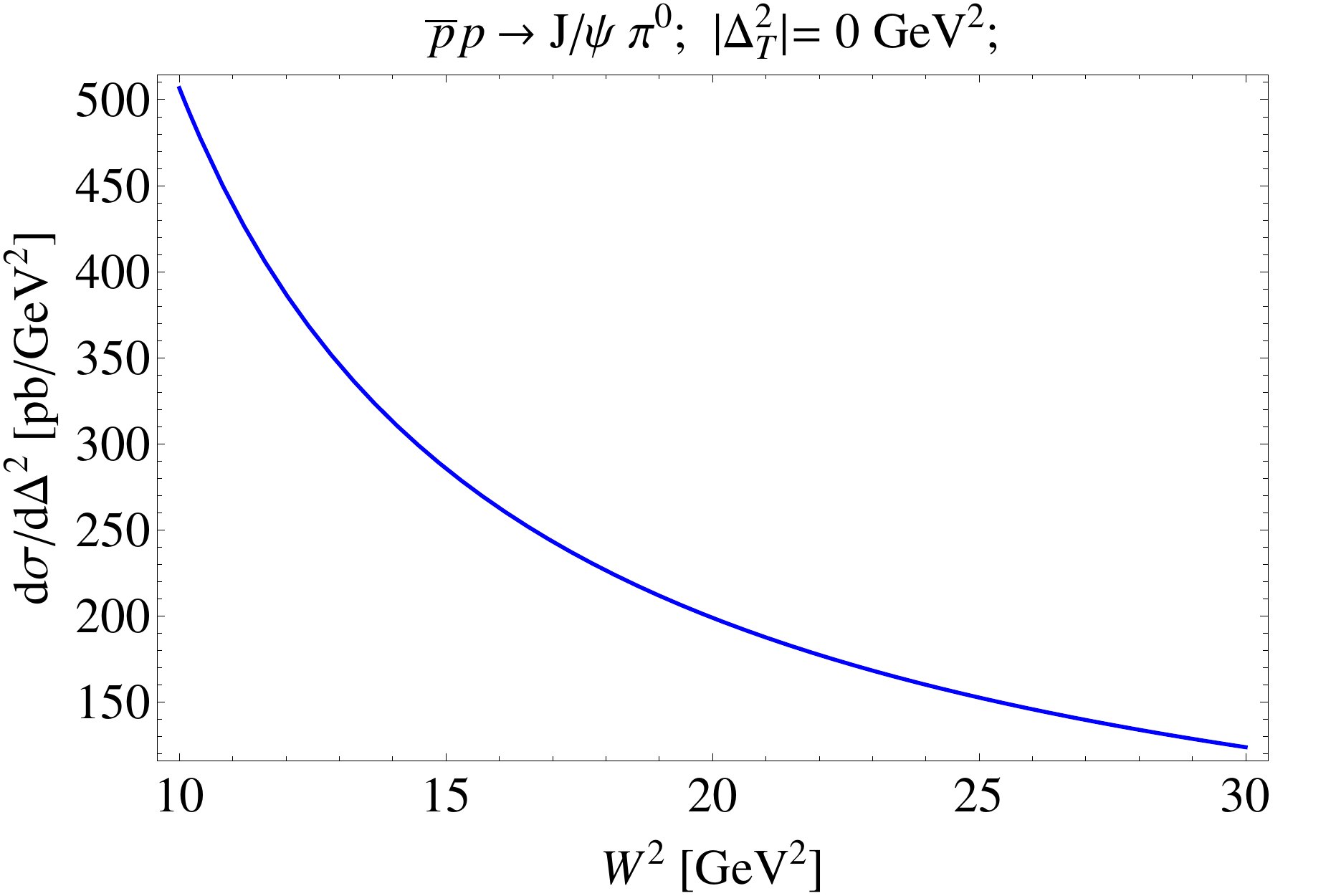} \ \ \  %~~~~~~~~~~~~~~~
  \includegraphics[width=0.45\textwidth]{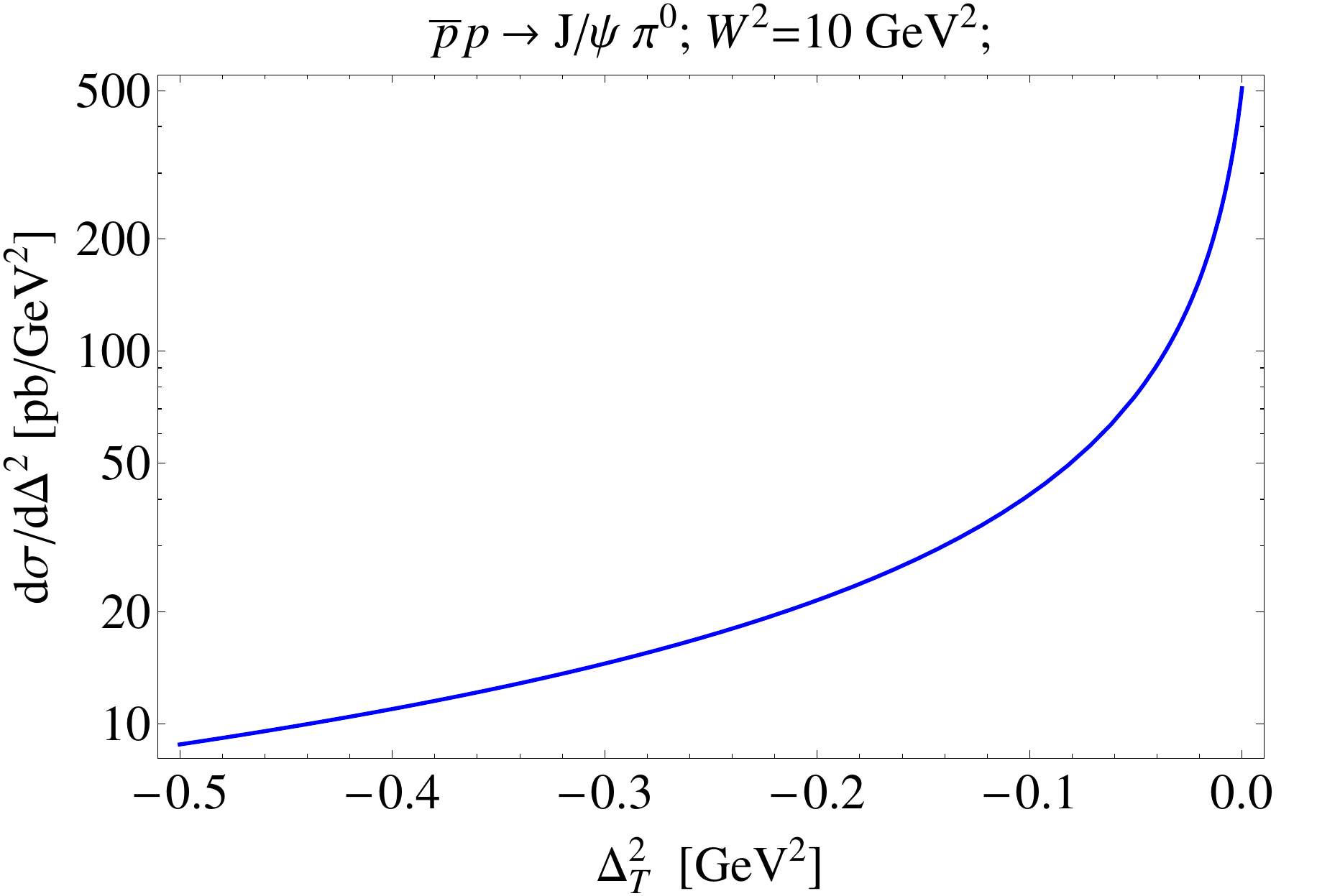}
  \caption{{\bf Left:} Cross section
$\frac{d \sigma}{d \Delta^2}$
for
$p \bar{p} \to J/\psi \pi^0$ as a function of $W^2$ for $\Delta^2_T=0$;
 {\bf Right:}  Same cross section for $W^2=10$ GeV$^2$ as a function
 of the transverse momentum transfer squared $\Delta^2_T$.}
 \label{CS_JPsi_at_PANDA}
  \end{center}
\end{figure}

The  mechanism involving nucleon-to-meson TDAs was also proposed in Ref.~\cite{PsiTDA}
for the reaction
\begin{equation}
  N (p_N) \;+ \bar N (p_{\bar N}) \; \to  J/\psi(p_{\psi})\;+\; {\cal M}(p_{{\cal M}}),
  \label{reac_Jpsi}
\end{equation}
which  can also be studied at \={P}ANDA alongside with the investigation
of the spectrum of charmonium states. Similarly to
(\ref{BarNNannihilation reaction}),
one can consider the near-forward (see the right panel of Fig.~\ref{Fig_TDAS_at_PANDA})
and the near-backward kinematic regimes
symmetric due to  charge conjugation invariance.
On Fig.~\ref{CS_JPsi_at_PANDA} we show the differential cross section
$\frac{d \sigma}{d \Delta^2}$
for
$p \bar{p} \to J/\psi \,\pi^0$
as a function of $W^2$ for exactly forward (backward) pion production ($\Delta^2_T=0$)
and as a function of $\Delta_T^2$ for a fixed value of $W^2$.
The cross channel nucleon exchange model
(\ref{Nucleon_exchange_contr_VAT})
is employed for the relevant $\pi N$ TDAs. To cope with the strong
$\sim \alpha_s^6$
dependence of the cross section we fix the value of $\alpha_s$
from the requirement that it allowed to reproduce the experimental value
of
$\Gamma(J/\psi \to p \bar{p})$
within the perturbative QCD description of Ref~\cite{Chernyak:1987nv}.
Based on the cross-section estimates presented in Ref.~\cite{PsiTDA}
a detailed feasibility study for
accessing
$p \bar{p} \to J/\psi \pi^0$
at  \={P}ANDA
has been performed
\cite{Singh:2014pfv, Ma, Singh:2016qjg}.

\section{Conclusion}
Baryon-to-meson TDAs are new non-perturbative objects which have been
designed to help us scrutinize the inner structure of nucleons.
Experimentally, one may access  TDAs both in the space-like domain
with backward electroproduction of mesons at JLab and COMPASS
and in the time-like domain in antiproton nucleon annihilation at \={P}ANDA.
We also expect
\cite{JParc}
the time-reversed
$\pi \to N$
TDAs to be accessible at J-Parc  through the reactions:
$\pi^- p \to J/\psi \,n\,$;
$\pi^- p \to \gamma^* \,n \to  \mu^+ \mu^- n$.
Extracting TDAs from space-like and time-like reactions will be a stringent test of their universality
\cite{Muller:2012yq}, and hence of the
factorization property of hard exclusive amplitudes. This hopefully will help us to disentangle details of the complex dynamics of quark and gluon confinement
in hadrons.

The special  
$\pi N$  TDA 
case  does not exhaust all interesting possibilities, and the vector meson 
sector should be experimentally accessible as well as the pseudoscalar meson sector
\cite{Pire:2015kxa}.
A  double handbag description of other processes such as charm meson pair production in proton-antiproton annihilation may also necessitate the introduction of baryon to charmed meson TDAs
\cite{Goritschnig:2012vs}.

\begin{acknowledgements}
This work is partly supported by grant No 2015/17/B/ST2/01838 by the National Science Center in Poland, by the French grant ANR PARTONS (Grant No. ANR-12-MONU-0008-01), by the COPIN-IN2P3 agreement, by the Labex P2IO and by the Polish-French collaboration agreement Polonium.
K.S. acknowledges the support from the Russian Science Foundation (Grant No. 14-22-00281).
\end{acknowledgements}

% BibTeX users please use
%\bibliographystyle{spbasic}


\end{document}